\documentclass{ws-procs9x6}
\begin{document}
\title{THE ROLE OF VARIATIONS OF CENTRAL DENSITY OF WHITE DWARF PROGENITORS UPON TYPE IA SUPERNOVAE}
\author{R. FISHER$^*$, D. FALTA$^*$, G. JORDAN$^{**}$, and D. LAMB$^{**}$}
\address{$^*$ Department of Physics, University of Massachusetts Dartmouth,\\
North Dartmouth, MA 02747-2300, United States\\
$^{**}$Department of Astronomy \& Astrophysics, University of Chicago,\\
Chicago, IL.  60637, United States \\
E-mail: robert.fisher@umassd.edu\\
Web : http://www.novastella.org}
\begin{abstract}
The discovery of the accelerated expansion of the universe using Type
Ia supernovae (SNe Ia) has stimulated a tremendous amount of interest
in the use of SNe Type Ia events as standard cosmological candles, and
as a probe of the fundamental physics of dark energy. Recent
observations of SNe Ia have indicated a significant population
difference depending on the host galaxy. These observational findings
are consistent with SNe Ia Ni-56 production in star-forming spiral
galaxies  some 0.1 solar masses higher -- and therefore more luminous
-- than in elliptical galaxies. We present recent
full-star, 3D simulations of Type Ia supernovae which may help
explain the nature of this systematic variation in SNe Ia
luminosities, as well as the nature of the Ia explosion mechanism.
These insights may in turn eventually shed light on the mystery of
dark energy itself.
\end{abstract}
\keywords{Type Ia supernovae; computational astrophysics.}
\bodymatter
\section{Introduction : Observational Properties of Type Ia Supernovae}

The discovery of the accelerated expansion of the universe using Type Ia supernovae \cite{riessetal98, perlmutteretal99} has stimulated a tremendous amount of interest in the use of SNe Type Ia events as standard cosmological candles, and as a probe of the fundamental physics of dark energy.

Supernovae come in different types, classified by their spectra. They are grouped into two main categories based on the presence or absence of Balmer absorption lines, indicating the presence or absence of atomic hydrogen. Type II contain hydrogen, whereas Type I lack hydrogen. These types are further subdivided by absorption line features and light curve shapes, see Table \ref{table1}. In particular, Type Ia exhibit an absorption line of singly ionized silicon (Si$^+$) at 615 nm at peak light. Type Ib/c are further classified according to the absence of Si II lines, and the presence or absence of He lines, respectively. 

Accordingly, Type Ia properties are consistent with white dwarf progenitors. The two leading models include white dwarfs in binary systems -- either with a non-degenerate companion, the so-called ``single-degenerate'' scenario, or with another white dwarf -- the so-called ``double-degenerate'' scenario. While both types of events are likely to occur in nature, and contribute to the diversity of explosion energies, there is still considerable debate over their relative frequencies, and even what the explosion mechanism is in each case. We outline some key theoretical and mounting observational evidence below based on nearby historic supernovae, which supports the identification of typical brightness ``Branch normal'' supernovae with single-degenerate events.
\begin{table}
\tbl{Classification of Supernovae}
{\begin{tabular}{ c c c }
Mass $\longrightarrow$ & & $\longleftarrow$ Age \\
\toprule
SN Ia & SN II (IIn, IIL, IIP, IIb) & SN Ib/c \\
\colrule
No Hydrogen & Has Hydrogen & No Hydrogen \\
(Si$^+$ absorption) & & (no  Si$^+$) \\
White dwarf Progenitor & Core collapse of & Core collapse \\
 & a massive star & (outer layers stripped by winds) \\
\botrule
\end{tabular}}
\label{table1}
\end{table}

Tying the variable luminosities of Type Ia explosions to additional parameters is of key significance in any attempt to calibrate the Phillips relation. Analysis of light spectra demonstrates that Type Ia supernovae produce a relatively constant combined yield of stable nuclear statistical equilibrium  nuclei (NSE)  of  $^{58}$Ni and $^{54}$Fe and intermediate mass elements  (IME) Si-Ca of $1.05 \pm .09\ M_{\odot}$ \cite{mazzalietal07}. The approximately fixed burned mass over a wide range of SNe Ia luminosities suggests that SNe Ia share a single common explosion mechanism. However, recent observations of SNe Type Ia have indicated a significant population difference depending on the host galaxy. In particular, observations have found that SNe Ia in star-forming galaxies decline more slowly than those in ellipticals \cite{howelletal07}.  These findings are consistent with   Ia SNe $^{56}$Ni production  in star-forming galaxies  some $\sim 0.1\ M_{\odot}$ higher than that in ellipticals \cite{pirobildsten08}. 

Previous work has focused on the possible influence of the metallicity as a second parameter in explaining the luminosity variance of Type Ia events \cite{timmesetal03, pirobildsten08, chamulaketal08, mengetal08}. These results suggest that while the metallicity effect may indeed contribute to the observed variance in Type Ia luminosities, alone it accounts for only a portion of the total variance \cite {timmesetal03, pirobildsten08}. Consequently, we must look elsewhere to other physical effects -- including both the central  density and angular momentum profile of the white dwarf progenitors -- which may explain the majority of the observed Ia luminosity variance. These progenitor properties may in turn be directly influenced by the surrounding environment of Ia event, in particular the accretion rate from the companion star. 

To put the single- and double-degenerate white dwarf progenitor models into their proper astrophysical context, we briefly consider the observational evidence. A number of recent papers have lended strong support for the general viability of the single-degenerate channel for the origin of Branch normal SNe Type Ia, in which a progenitor white dwarf accretes material from  a non-degenerate companion star \cite{whelaniben73, nomoto82}. Analysis of the spectrum from the Type Ia SN 2006X in Virgo suggests that the blast wave moved through the circumstellar medium and collided with the ejecta from an otherwise undetected red giant companion star \cite{patatetal07}. Furthermore,  a candidate G-type companion to SN1572 (Tycho)  has been identified \cite{ruizlapuenteetal04}. SN1572 has independently been identified as a  Branch-normal Ia event  based on analysis of historical light curves and light echo spectra \cite{ruizlapuente04, krauseetal08}. 

However, despite this string of recent successes, the single degenerate model faces significant theoretical challenges. In particular, models suggest that a non-rotating white dwarf can burn stably only over a relatively narrow mass accretion rate range \cite{nomoto82}. The addition of differential rotation broadens this range \cite{yoonlanger04}. However, when one introduces rotation to the white dwarf structure, one must make a number of assumptions regarding accretion and internal shear in order to understand the evolution of the angular momentum distribution of the white dwarf progenitor. The accretion rate in turn sets the central density of the white dwarf at ignition.

As a consequence of these challenges faced by theoretical descriptions of the single degenerate model, and the limited direct observational evidence constraining progenitors, there is a significant degree of uncertainty in the characterization of the degenerate progenitor.  Here we focus on the role which the central density (or equivalently, the total mass) of a non-rotating white dwarf progenitor plays in determining the luminosity of Type Ia SNe.

\section{Physics of Type Ia Supernovae}

The energetics of the single-degenerate model of Type Ia supernovae can be simply estimated using nothing more than elementary physics. The internal energy of $N$ electrons in a fully-degenerate white dwarf is $N$ times the characteristic Fermi energy $E_F$.  The Fermi energy can itself be estimated for a relativistic electron with momentum $p$ as
\begin {equation}
E_F = p c
\end {equation}
Applying the Heisenberg Uncertainty Principle, $p \sim \hbar n^{\frac{1}{3}}$, where $n$ is the number density of electrons. Therefore, 
\begin{equation}
E_F \sim \hbar n^{\frac{1}{3}}c \sim \frac{\hbar N^{\frac{1}{3}}c}{R}
\end{equation}
Where in the second step we have estimated the mean number density within a white dwarf of radius $R$.
Including the gravitational binding energy, the white dwarf has a total energy of
\begin{equation}
E = NE_F + E_G \sim \frac{\hbar N^{\frac{4}{3}}c}{R} - \frac{GN^2m_p^2}{R}
\end{equation}
here $E_G$ is the binding energy, and $m_p$ is the proton mass. Note that we focus here on the essential physics, and have therefore suppressed all factors of order unity, and neglected composition effects.

This elementary analysis reveals a remarkable feature of a fully degenerate star. In a fully degenerate gas, the  degeneracy pressure is fundamentally independent of temperature. This immediately leads to the result that both the internal energy term and the gravitational binding energy term scale inversely with radius. Consequently, a fully-degenerate white dwarf cannot seek a lower total energy state by an adiabatic spherical compression or expansion. This stands in sharp contrast to stars supported by ordinary gas pressure, which can indeed minimize their total energy by spherical adiabatic compression or expansion. We are led to conclude that the stability of the white dwarf is therefore set solely by the sign of the total energy $E$. The critical case is where the total energy $E$ equals zero; solving for the maximum total mass $M_{\rm Chandra}$ of the star then yields   
\begin{equation}
M_{\rm Chandra} \sim \left( \frac{\hbar c}{G} \right)^{\frac{3}{2}} \frac{1}{m_p^2} \sim m_{\rm Planck} \left( {m_{\rm Planck} \over {m_p} }\right)^2 \sim 1.5\ M_{\odot} 
\end{equation}
This analysis shows the critical mass, which is known as the Chandrasekhar limit, to be a combination of fundamental physical constants. Indeed, it is fundamentally set by the Planck mass times a large dimensionless number, which is the ratio of the Planck mass to the proton mass, squared.  This simple analysis, which neglects composition effects, demonstrates the Chandrasekhar mass is of order a solar mass; a more precise calculation demonstrates it to be $1.4\ M_{\odot}$ for a predominantly C/O white dwarf.

We take the progenitor to be a carbon-oxygen white dwarf. The mass, which is close to the Chandrasekhar limit,  undergoes carbon burning, releasing roughly $10^{18}$ ergs/g. The characteristic nuclear energy available for a Type Ia is about $3 \cdot 10^{51}$ ergs $\sim 3$ foe, where 1 foe is a convenient unit representing $10^{51}$ ergs, a typical Type Ia luminosity. Therefore, there is more than sufficient nuclear energy within the white dwarf progenitor to represent not only typical Type Ia events, but also the more luminous events observed. 

In the single-degenerate model of a Type Ia supernova event, the system begins as a binary pair of main sequence stars. As time progresses, the more massive star evolves into a giant, accreting gas onto its companion, which expands and is eventually engulfed. The core of the giant and its companion begin to spiral inward inside a common envelope.  Tidal torques cause the envelope to be ejected, and the binary separation to decrease. The core of the giant then collapses into a white dwarf that begins to accrete gas from its aging companion. The white dwarf continues to accrete until reaching a critical mass close to but not equal to the Chandrasekhar limit. At a critical mass set by the accretion rate from the companion, the central core of the white dwarf ignites a nuclear flame that causes the  supernova explosion, and ejects the companion from the system.

The precise initial conditions leading to ignition are still poorly understood. The problem arises because of a large dynamic range between the long convection phase (on the order of a hundred years), and the brief deflagration/detonation phase following flame ignition, which is on the order of one to several seconds \cite{nomotoetal84}. At some point during convection, a runaway nuclear burning occurs during which the burning exceeds the neutrino cooling. The runaway reaction ignites a deflagration flame slightly off-center from the white dwarf, giving rise to a buoyancy force. This buoyant flame bubble undergoes subsonic burning while rising toward the surface. The possibility of the bubble burning though the entire star in a pure deflagration \cite{niemeyeretal96} has generally lost support because it tends to burn inefficiently, leaving behind a significant amount of nuclear fuel, and producing events which are generally underluminous with respect to typical Branch normal Ia events. It is most likely that the bubble either undergoes a deflagration to detonation transition (DDT) \cite{khokhlov91} or leads to a gravitationally-confined detonation (GCD) \cite{plewaetal04}.  In the case of a GCD, the deflagration bubble breaks through surface of the star. Ash is launched into the atmosphere and wraps around the surface of the star under the force of gravity. The ash quickly reaches the opposite side of the star and collides with itself. This collision creates a supersonic detonation front that unbinds the star.

At present, both the DDT and GCD mechanisms have their advantages and disadvantages. The DDT model can be tuned to be consistent with observations. However, in the absence of a full first-principles understanding of the detonation transition, DDT simulations set the detonation transition as a parameter. The current belief is that detonation can occur when the flame is ripped apart by turbulence, and transitions into the distributed burning regime.  However, the precise conditions under which this transition occurs remain a matter of intense research, so that the transition density is largely a free parameter in the simulations.  On the other hand, the GCD mechanism can successfully produce detonations without fine-tuning, though initial models typically overproduced $^{56}$Ni, underproduced intermediate mass elements (IME), and generated  overluminuous events. We address one aspect of the luminosity problem in the following section, by varying the central density of the white dwarf progenitor models in the simulations.

\section{Simulations of Type Ia Supernovae}
We begin with a carbon-oxygen white dwarf with a pre-ignited flame bubble of initialized size and location. The model is run through  detonation. The evolution of the white dwarf and bubble has a broad range of length scales. On the largest scales, one must be able to follow the expansion of the supernovae out to several tens of thousands of kilometers to capture the homologous expansion phase, and on the smallest scales, one must begin to capture the flame physics, which extends down to the laminar flame thickness on centimeter scales. We employ the Paramesh library within the FLASH code to implement an adaptive mesh refinement (AMR) mesh. Even the power of AMR still does not allow one to avoid the enormous dynamic range between the large-scale physics of the explosion and the flame scale -- over $10^9$ in linear dimension -- so we incorporate a thickened model of the flame surface, which artificially thickens the flame over $\sim 4$ grid cells.

We evolve the simulation using a comprehensive multiphysics model, including the coupled equations of hydrodynamics, self-gravity, and nuclear combustion. Specifically, we incorporate the Euler equations of hydrodynamics :
\begin{eqnarray}
&&\frac{\partial \rho}{\partial t} + \vec{\nabla} \cdot (\vec{v} \rho) = 0 \\
&&\frac{\partial \rho \vec{v}}{\partial t} + \vec{\nabla} \cdot (\vec{v} \otimes \rho \vec{v}) = - \vec{\nabla} P - \rho \vec{\nabla} \Phi \\
&&\frac{\partial \rho E}{\partial t} + \vec{\nabla} \cdot [\vec{v} (\rho E + P )] = \rho \vec{v} \cdot \vec{\nabla}\Phi + \rho \epsilon_{nuc}
\end{eqnarray}
with source terms for self-gravity and nuclear energy release. Here $\rho$ is mass density, $\vec {v}$ is velocity, $P$ is pressure, $\Phi$ is gravitational potential, $\epsilon_{nuc}$ is the specific nuclear energy release.  $E$ is the total energy density,  given by
\begin{equation}
E = \rho \left (U + \frac{1}{2} v^2 \right)
\end{equation}
Where $U$ is the specific internal energy. 

The Euler equations are coupled to Poisson's equation for self-gravity
\begin{equation}
{\nabla}^2\Phi = 4\pi G \rho
\end{equation}
and an advection-diffusion reaction model of the thickened combustion front
\begin{eqnarray}
&&\frac{\partial \phi}{\partial t} + \vec{v} \cdot \vec{\nabla}\phi = \kappa {\nabla}^2 \phi + \frac{1}{\tau}R(\phi) \\
&&R(\phi) = \frac{f}{4} (\phi - \epsilon_0)(1-\phi + \epsilon_1)
\end{eqnarray}
Here $\phi$ is a scalar progress variable which monitors the advancement of the flame surface, and sets the nuclear energy release function $\epsilon_{\rm nuc}$ above. $\kappa$ is a parameter representing the diffusivity of the thickened flame, $\tau$ is the reaction timescale, $R$ is the reaction term, which is set by a lowered Kolmogorov-Petrovski-Piscounov (KPP) binomial, specified by the three parameters,  $f$, $\epsilon_0$, and $\epsilon_1$.

\begin{figure}
\begin{center}
\includegraphics [width=0.75 \textwidth] {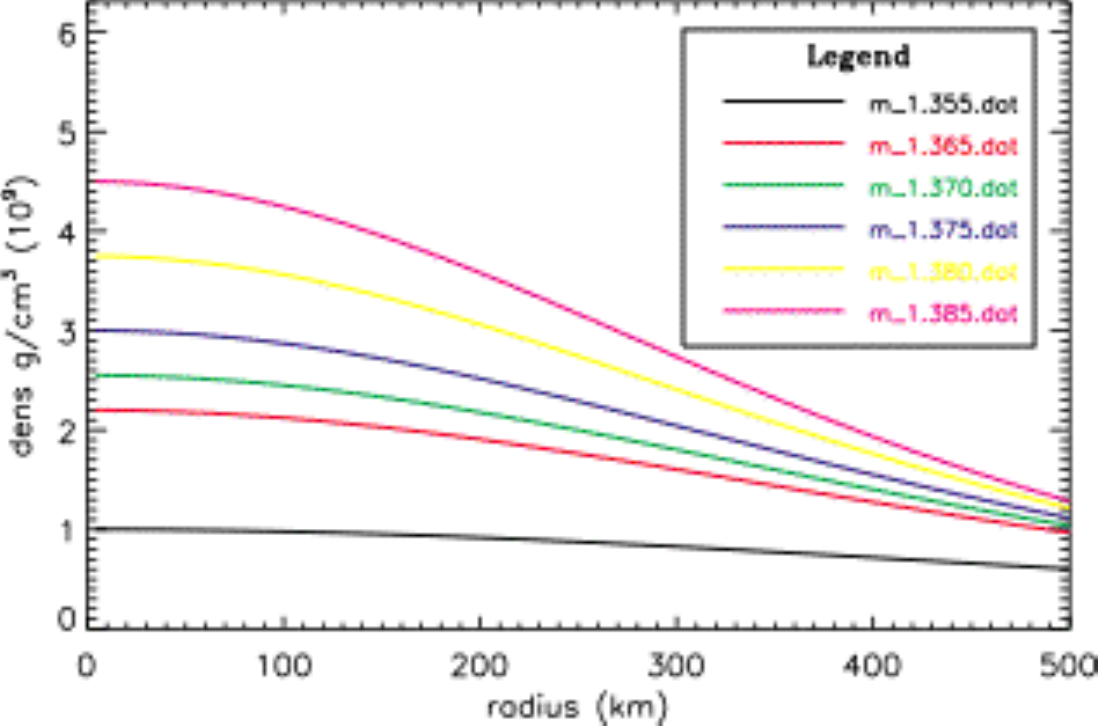} 
\end{center}
\caption {Density versus radius for a series of cold white dwarf models, ranging from $1.355\ M_{\odot}$ to $1.385\ M_{\odot}$. The plot focuses on the innermost region of the white dwarf to emphasize the sensitive dependence of central density on total mass.}
\label{centraldensity}
\end{figure}

Our first sucessful 3-D simulation of a Type Ia detonation was a cold white dwarf model in initial equilibrium with initial mass 1.36 $M_{\odot}$. The nuclear bubble was ignited within a spherical region slightly offset from the center of the white dwarf. The supercomputer simulation ``marches" this condition forward in time in 3-D, using full equations describing the flame, hydrodynamics, and self-gravity.
An numerical analysis yields a critical conditions for initiation \cite{niemeyeretal97}. We found that the critical conditions for degenerate white dwarf matter are robustly met in our 3-D simulations. We have confirmed that detonation arises independent of the resolution in the detonation region, and also for a wide variety of initial bubble sizes and offsets. Simulations of the GCD model produce intermediate mass elements at a velocity coordinate $\sim 11,000$ km/s, creating a layered structure of IME and Fe peak (NSE) products similar to observation \cite{mazzalietal07}.

However, despite these successes, initial simulated 3D GCD models generally underproduced intermediate mass elements and overproduced Ni. Consequently, the models are generally too luminous in comparison to Branch normal Ia events. The underabundance of IME suggest that additional pre-expansion is required to produce Ni and IME abundances consistent with observation. Moreover, other 3D simulations were found to pre-expand significantly and lead to failed explosions \cite{ropkewoosley06}. 

One factor influencing this outcome is the uncertainty in the central density of the progenitor model, which is fundamentally set by the accretion rate in the single degenerate scenario. As shown in figure 1, a small change in the total mass of a cold 50/50 C/O white dwarf progenitor model from $1.355\ M_{\odot}$ to $1.385\ M_{\odot}$ leads to more than a factor of four change in central density : from $10^9$ gm/cm$^3$ to $4.4 \cdot 10^9$ gm/cm$^3$.  This increase in central density in turn directly impacts the nuclear energy release during the deflagration phase. Fundamentally, this is due to two reasons. First,  the critical wavelength for Rayleigh-Taylor instability $\lambda_c \propto S_l^2 / g$ is sensitively dependent on the density through the dependence on the laminar speed $S_l$. Here $g$ is the gravitational acceleration. Higher central densities lead to a larger critical wavelength for the development of the Rayleigh-Taylor instability, which in turn suppresses the growth of turbulence during the early stages of the burning, and leads to an enhancement in the burnt mass. Second, the increased total mass of the progenitor with increased central density causes it to be closer to the Chandrasekhar mass, and therefore more easily pre-expanded than models with lower central density.

The net effect on the energy release during the deflagration phase in the GCD model is dramatic, as shown in figure 2. An increase in total mass by just one percent, and an increase in central density by a factor of two, leads to a four-fold increase in the total fractional energy release during the deflagration phase. This leads to a significantly enhanced pre-expansion rate in the case of the higher central density model.

\begin{figure}
\begin{center}
\includegraphics [width=0.75 \textwidth] {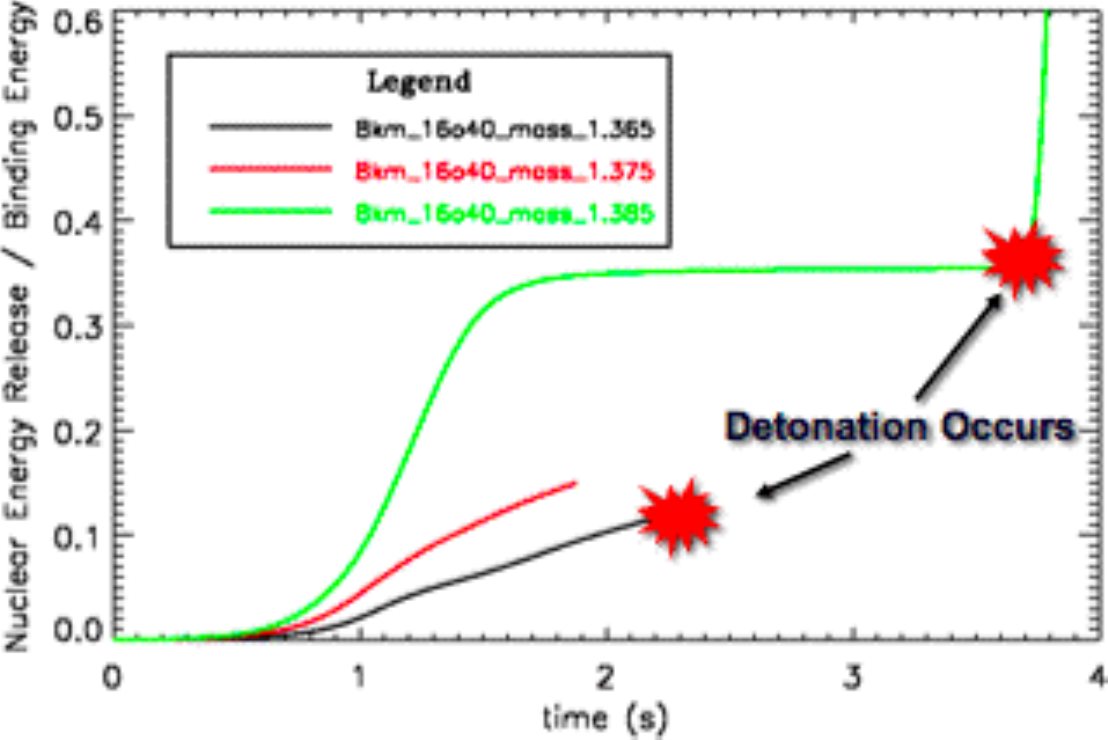} 
\end{center}
\caption {A plot of the nuclear energy released as a function of time,  shown for a range of stellar masses, from 1.365 $M_{\odot}$ to 1.385 $M_{\odot}$. }
\end{figure}

\section{Conclusion}
The GCD model simulations demonstrate that successful self-consistent detonations of Type Ia supernovae in 3-D are possible without having to be initiated by hand.  The nuclear energy release during the deflagration phase is found to be sensitively dependent upon the central density of the white dwarf progenitor model.  With a greater pre-expansion afforded by the greater nuclear energy release in higher central density models, the GCD mechanism can lead to observed Branch normal luminosities, and typical levels of intermediate mass elements (Si -Ca). 

\section*{Acknowledgments}
The software used in this work was in part developed by the DOE-supported ASC / Alliance Center for Astrophysical Thermonuclear Flashes at the University of Chicago.
The PARAMESH software used in this work was developed at the NASA Goddard Space Flight Center and Drexel University under NASA's HPCC and ESTO/CT projects and under grant NNG04GP79G from the NASA/AISR project.

\end{document}